\documentclass[sigconf]{acmart}

\usepackage{booktabs} % For formal tables
\usepackage{amssymb}
\usepackage{graphicx}
\usepackage{algorithm}
\usepackage{algorithmic}
\usepackage{multirow}
\usepackage{makecell}
\usepackage{url}

\setcopyright{rightsretained}
\acmConference[Social Web in Emergency and Disaster Management]{ACM WSDM conference}
\acmYear{2018}
\begin{document}
\title{Unsupervised Hashtag Retrieval and Visualization for Crisis Informatics}
% \titlenote{Produces the permission block, and
% \subtitle{Extended Abstract}
% \subtitlenote{The full version of the author's guide is available as
%   \texttt{acmart.pdf} document}

\author{Yao Gu}
% \authornote{The secretary disavows any knowledge of this author's actions.}
\affiliation{%
  \institution{Department of Computer Science}
  \streetaddress{University of Southern California}
  \city{Los Angeles} 
  \state{California} 
  \postcode{90089}
}
\email{yaogu@usc.edu}

\author{Mayank Kejriwal}
\affiliation{%
  \institution{Information Sciences Institute}
  \streetaddress{USC Viterbi School of Engineering}
  \city{Marina Del Rey} 
  \state{California} 
  \postcode{90292}
}
\email{kejriwal@isi.edu}

% \authornote{This author is the
%   one who did all the really hard work.}
% \affiliation{%

% \affiliation{%
% \affiliation{%
% \affiliation{%
%   \country{China}
% }

% \affiliation{%


% \author{Julius P.~Kumquat}

\renewcommand{\shortauthors}{Gu and Kejriwal}

\begin{abstract}
In social media like Twitter, hashtags carry a lot of semantic information and can be easily distinguished from the main text. Exploring and visualizing the space of hashtags in a meaningful way can offer important insights into a dataset, especially in crisis situations. In this demo, we present a functioning prototype, HashViz, that ingests a corpus of tweets collected in the aftermath of a crisis situation (such as the Las Vegas shootings) and uses the fastText bag-of-tricks semantic embedding algorithm (from Facebook Research) to embed words and hashtags into a vector space. Hashtag vectors obtained in this way can be visualized using the t-SNE dimensionality reduction algorithm in 2D. Although multiple Twitter visualization platforms exist, HashViz is distinguished by being simple, scalable, interactive and portable enough to be deployed on a server for million-tweet corpora collected in the aftermath of \emph{arbitrary} disasters, without \emph{special-purpose} installation, technical expertise, manual supervision or costly software or infrastructure investment. Although simple, we show that HashViz offers an intuitive way to summarize, and gain insight into, a developing crisis situation. HashViz is also completely unsupervised, requiring no manual inputs to go from a raw corpus to a visualization and search interface. Using the recent Las Vegas mass shooting massacre as a case study, we illustrate the potential of HashViz using only a web browser on the client side. 
\end{abstract}




\maketitle

% call-for-papers URL: https://sites.google.com/site/swdm2018/call-for-papers
% demo papers should be 2 pages

\section{Demo}
\begin{figure}
\centering
\includegraphics[height=2.5in, width=3.5in]{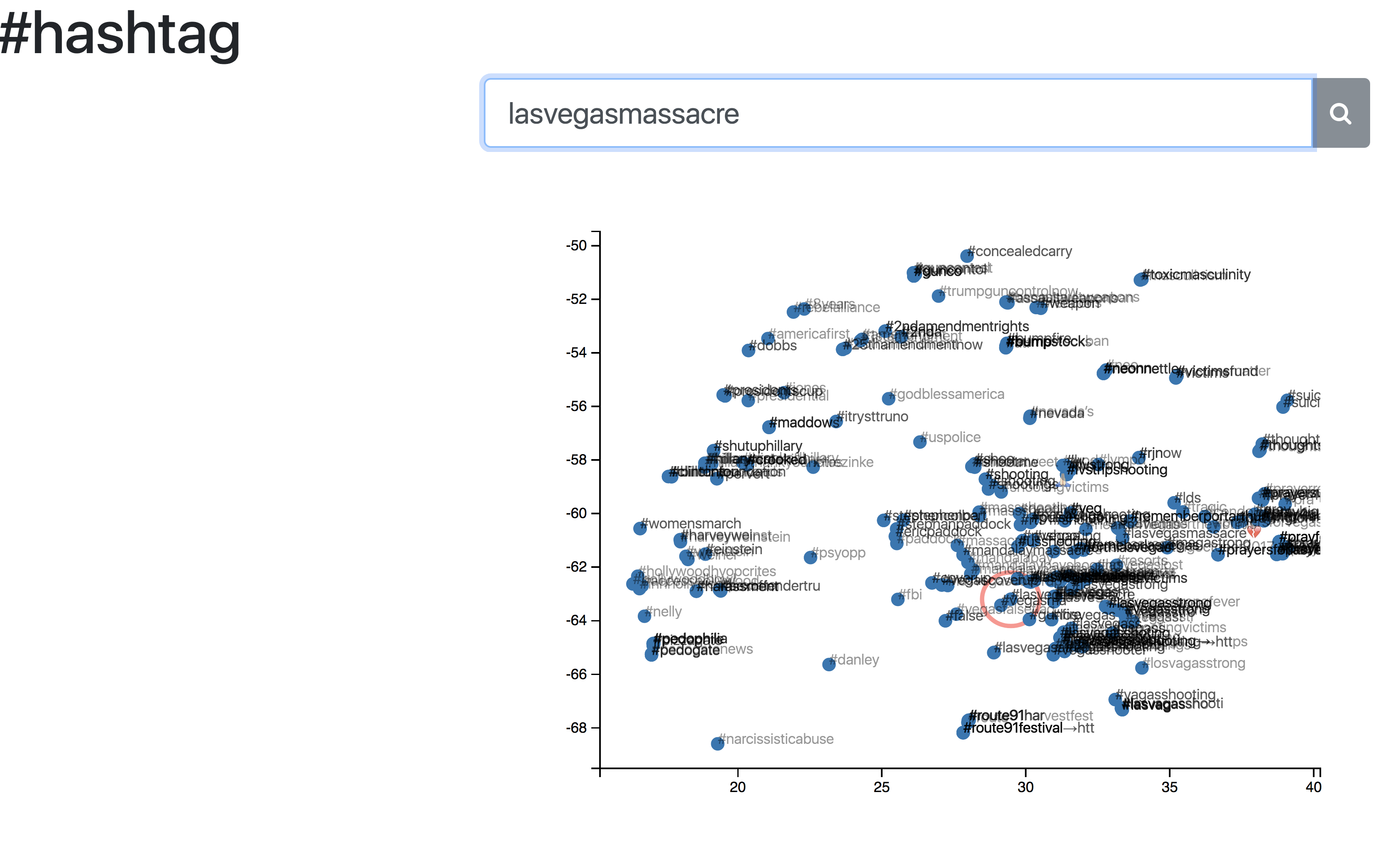}
\caption{Display after searching for hashtag `lasvegasmassacre'.}\label{fig1}
\end{figure}

\begin{figure}
\centering
\includegraphics[height=2.8in, width=3.5in]{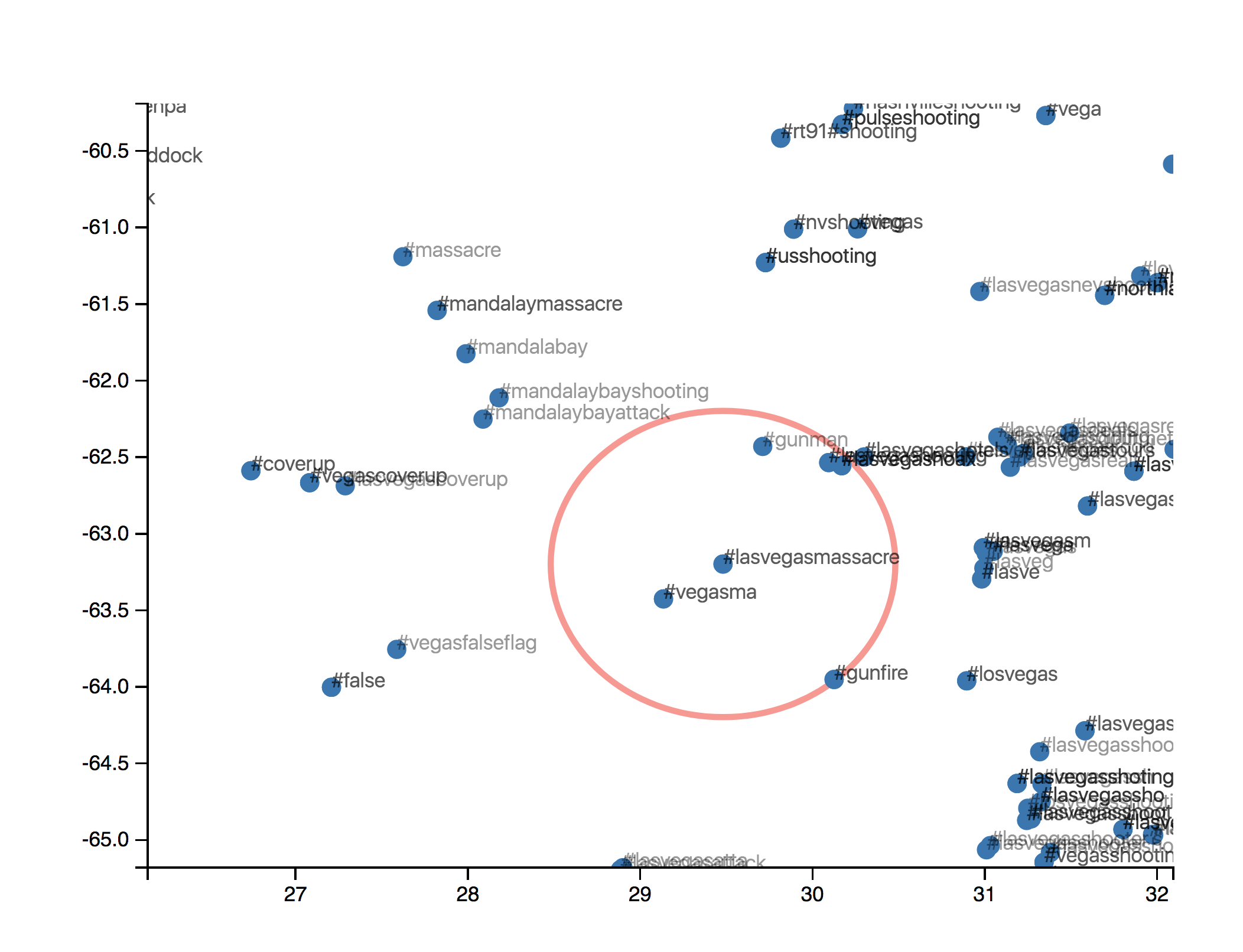}
\caption{Display after drawing a bounding box around `lasvegasmassacre' and zooming in.}\label{fig2}
\end{figure}

\begin{figure}
\centering
\includegraphics[height=2.8in, width=3.5in]{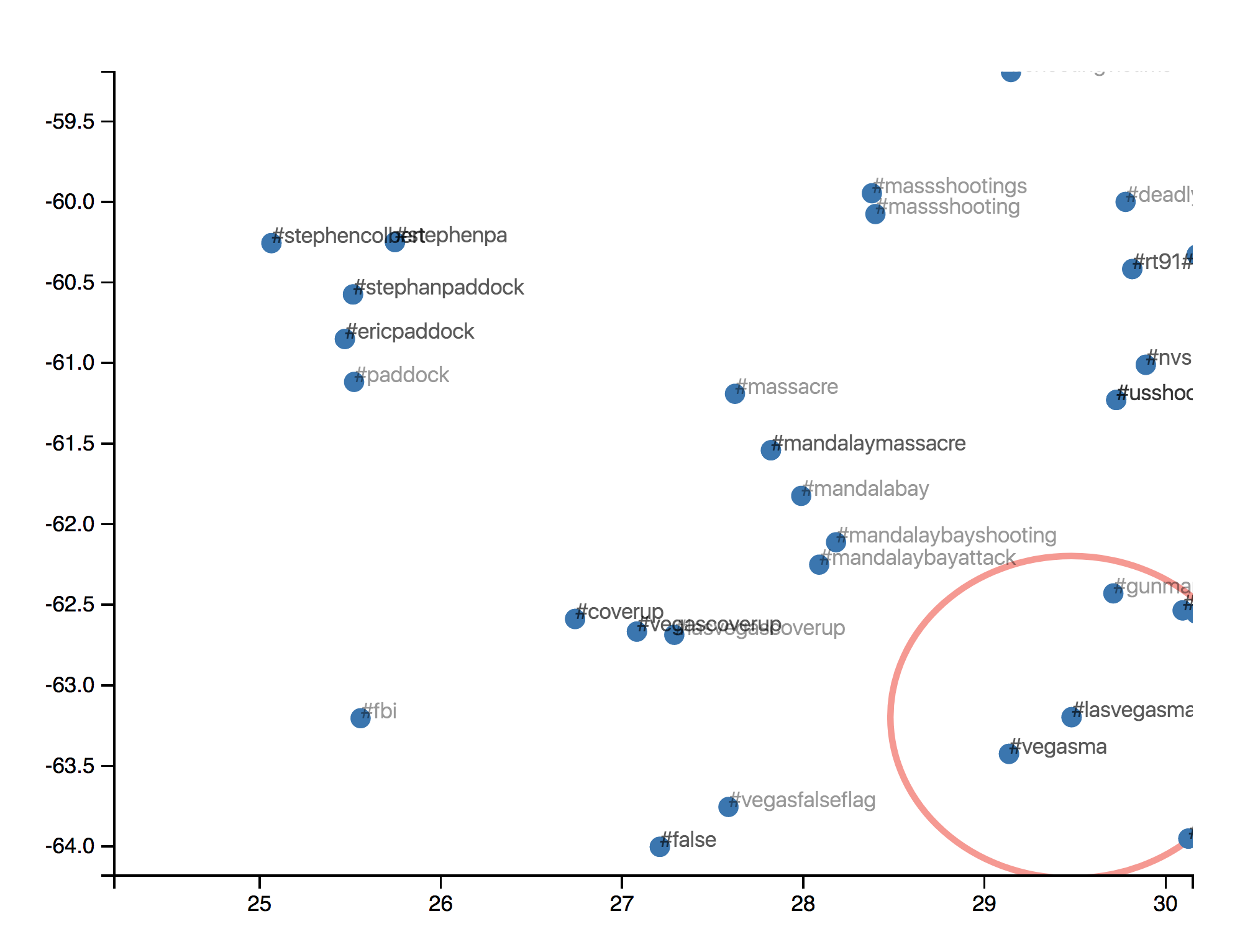}
\caption{Display allows a user to interactively scroll around the plot.}\label{fig3}
\end{figure}

We provide a high-level overview of the demo via the screenshots in Figures \ref{fig1}, \ref{fig2} and \ref{fig3}. Following pre-computation steps such as corpus preprocessing, embedding and t-SNE visualization, the user may commence the demo by entering a hashtag query such as `lasvegasmassacre'. Using the semantic text embeddings, the system, in real-time, computes the hundred nearest neighbors of the hashtag (if it exists\footnote{Currently, the system does not work for non-existing hashtags, but we are working to extend its capabilities thereof by drawing on the generalization offered by the text embeddings.}) and returns an \emph{interactive} scatter plot. Thus, in an unsupervised way, and without relying on storage of original text or tweet data, the system is able to limit the display to only a hundred relevant entries, from tens of thousands of hashtags in the corpus.  Potential interactions (which may be interleaved) include drawing a bounding box and zooming into any region of the plot, and scrolling around the plot to explore the neighborhood.

In the scatter plot, the hashtag is encircled in red, and one can intuitively note the similarities between that hashtag and other hashtags in the corpus (Figure \ref{fig2}).  Similar hashtags, in the text embedding space (which uses the context to determine similarity) tend to be closer together, even when they share little in common in terms of the raw string. For example, when the user draws a bounding box around `lasvegasmassacre' and zooms in, \#lasvegasmassacre and \#mandalaybayattack appear relatively close, even though they share no terms in common. This has been achieved without any manual supervision or labeling. 

The user can also scroll around the `hashtag space' to gain situational awareness and insight that she might not fully be aware of otherwise. If we go a little bit further, in Figure \ref{fig3} for example, we gain an additional piece of information (the name of the shooter i.e. Stephen Paddock).

The entire process can be restarted simply by entering another query. The system is currently deployed on a cloud server and is completely unsupervised.

\section{Related Work}

Visualization is an important part of any human-centric system that is attempting to make sense of a large amount of information. Several good crisis informatics platforms that provide visualizations include Ushahidi \cite{ushahidi}, Twitris \cite{twitris}, Twitcident \cite{twitcident}, AIDR \cite{aidr}, CrisisTracker\cite{crisistracker}, TweetTracker \cite{tweettracker}, and several others \cite{choi2015real}, \cite{thom2015can}. Our preliminary system is meant to complement, not compete with, the capabilities offered by these platforms, since it provides succinct `summarization'-based visualizations that are unsupervised, can run on large and potentially noisy corpora, and that are specifically tuned to rapidly explore the space of Twitter \emph{hashtags} in crisis situations. 

In addition to this body of work, over the years, several sophisticated Twitter-specific visualization platforms have been proposed, but their complexity precludes them from rapid deployment and prototyping, especially in resource-constrained environments \cite{twitviz1}, \cite{twitviz2}, \cite{twitviz3}, \cite{twitviz4}. Some are specifically tuned for spatio-temporal analyses or community detection, and require access to text and other information. In contrast, our system is simple, scalable and portable, and relies on loosely coupled open-source components. Privacy concerns are also alleviated because hashtags, by themselves, do not carry sensitive information about individual users. The case study visualizations in Figures \ref{fig2} and \ref{fig3} show that hashtags can (even by themselves) provide useful situational and summarization-based insights for crisis informatics.  

In ongoing work, we are setting this system up so that it can be rapidly deployed on any Twitter corpus, possibly being updated in a streaming fashion, with only a few commands. We may also make the hashtags clickable, to access more insights related to that hashtag, and extend HashViz beyond just hashtags.

\bibliographystyle{ACM-Reference-Format}
\bibliography{sample-bibliography} 

\end{document}